# Global Dynamics in Galactic Triaxial Systems I

P. M. Cincotta,[1,2] C. M. Giordano[1,2] and M. J. Pérez[1,2]

[1] Facultad de Ciencias Astronómicas y Geofísicas, Universidad Nacional de La Plata, Paseo del Bosque S/N. La Plata (1900)-Argentina,

[2] Instituto de Astrofísica de La Plata (IALP), Consejo Nacional de Investigaciones Cientícas y Técnicas de la República Argentina.

**Abstract.** In this paper we present a theoretical analysis of the global dynamics in a triaxial galactic system using a 3D integrable Hamiltonian as a simple representation. We include a thorough discussion on the effect of adding a generic non–integrable perturbation to the global dynamics of the system. We adopt the triaxial Stäckel Hamiltonian as the integrable model and compute its resonance structure in order to understand its global dynamics when a perturbation is introduced. Also do we take profit of this example in order to provide a theoretical discussion about diffusive processes taking place in phase space.

## 1. Introduction

Observations with HST revealed the presence of very high stellar densities at the centers of early-type galaxies (Crane et al., 1993, Ferrarese et al., 1994), suggesting a power law ($r^{-\gamma}$) to fit them. The evidence for large central masses was also reforced from high-resolution kinematical studies of nuclear stars and gas, which disclosed the presence of compact dark objects with masses in the rage of $10^{6.5} - 10^{9.5} M_\odot$, presumably supermassive black holes (Ford et al., 1998). These observational results have produced a substantial change in the classic ideas on dynamics in triaxial galaxies.

Results obtained from numerical simulations show that the addition of a central mass to an integrable triaxial potential has deep effects on its dynamics, at least for the boxlike orbits which mainly cover the central region of triaxial galaxies. Black holes and central density cusps scatter these particular orbits during each close passage giving rise to chaos in the system. The sensitivity of boxlike orbits to deflexions also drives to a rounder central distribution of mass (see for instance Gerhard and Binney, 1985, and Udry and Pfenniger, 1988). This slow evolution towards axisymmetry suggests that stationary triaxial configurations could not exist for a central density cusp. Valluri and Merritt (1998) find that in most of early-type galaxies, the chaotic evolution would be determined by the mass of the central black hole, ($M_{bh}$), rather than by the slope of the density profile. They show that when the central mass contains 2% of the galaxy mass, a transition to global stochasticity sets up. For such large value of $M_{bh}$, the box–orbit phase space is almost completely stochastic and diffusive proccesses could take place in a very short timescales.

This result turned out to be substantially attractive because this critical black hole mass was close to the observed one (Kormendy and Ritchstone, 1995) and also close to the mass which induced a sudden evolution towards the axisymmetry in N–body simulations (Merritt and Quinlan, 1998). However, this is no longer true, since from the works of, for instance, Ferrarese and Merritt (2000) and Merritt and Ferrarese (2001) it has been known that the mass of black holes in galaxies from the black hole demographic relatioships are $0.1 - 0.2\%$ of the ellipsoid mass in which they are embedded.

Merritt and Fridman, (1996) arrive at the same conclusion analyzing two triaxial power law models: the steep ($\gamma = 2$) and the weak ($\gamma = 1$) cusp. They find, in agreement with Gerhard and Binney (1985), and Schwarzschild (1993), that triaxial galaxies with such huge concentration of mass would evolve towards a central axisymmetry, as box orbits gradually loose their distinguishability. For these models, in which a large fraction of phase space is dominated by a chaotic dynamics, the construction of self-consistent solutions requires the inclusion of stochastic orbits besides the regular ones. A system thus built evolves, mainly close to its center, as stochastic orbits mix through phase space. In order to obtain stationary solutions, they build the "fully-mixed models" keeping stochastic orbits out of the central part of the system where chaotic orbits mix ergodically driving to a rounder distribution and destroying the triaxial self-consistency. They find that though it is possible to build this kind of solutions for a weak cusp model, this is not the case for a strongly concentrated model. This would imply that triaxiality is not consistent with a high central density. The discussion about if nature is able to build

*Send offprint requests to*: María Josefa Pérez: e-mail: jperez@fcaglp.unlp.edu.ar

stationary non-axisymmetric stellar systems is still open. Let us recall however that Poon and Merritt (2002) were able to construct triaxial equilibria with central black holes which were both regular and stable, but under a very strong asumption. In fact, they need the chaotic building blocks being fully mixed for the triaxial equilibria to co-exist with central singularities, and there is no evidence yet that in 3D systems with divided phase space, a completely connected chaotic component actually exists. Moreover, it seems that this could happen only when the chaotic component has a large measure ($\sim 1$) and $t \to \infty$, which, from a physical point of view, would not be possible in galactic systems (see for instance Giordano and Cincotta, 2004 and reference therein and also Section 5).

All these conclusions rest on the strong hypothesis that diffusive processes drive stochastic orbits to mix, covering all chaotic region of phase space over a timescale of order of Hubble time. Taking into account the fundamental role that diffusive processes should play in galactic dynamics (as well as in asteroidal and planetary dynamics), we will discuss the relevance of diffusive processes in phase space theoretically as well as numerically. In this first paper we address the theoretical discussion while the numercal studies will be presented in a forthcoming paper.

Global dynamics of triaxial galactic systems were already studied in previous works, for example by Wachlin and Ferraz-Mello, (1998) and Papaphilippou and Laskar, (1998). They represented elliptical galaxies using, respectively, a generalization of a double-power-law spherical mass model and the axisymmetric softened version of the 3D logarithmic potential. For these models, they applied the frequency map analysis (Laskar, 1993) in order to investigate the orbital structure of the system. Both analysis revealed a strong indication of stochastic motion in both models. Though they show the main resonances of these systems and investigate the relevance of chaotic motion in these models, they did not focus on the study of the diffussive mechanisms which could be present in these systems. Moreover, the frequency map analysis is not an efficient tool for such kind of studies since it does not provide a good measure of chaos (see Cincotta and Simó, 2000, Cincotta et al., 2003).

In the present effort we first concentrate our study in computing the theoretical resonant structure of an integrable triaxial system. Particularly, we focus our attention to the Stäckel triaxial model, which we assume to be a rough approximation of an elliptical galaxy. Then we add a generic non–integrable perturbation in order to show how the resonace structure is distorted by the effect of this perturbation.

In section 2, we briefly recall the main characteristics of the integrable triaxial Stäckel model. Numerical details related to the computation of the resonances and the analysis of the resonace structure are presented in section 3. In section 4, we review and provide a theoretical discussion about the transition to chaos as a consequence of resonance interaction. Finally, we discuss the different processes that could lead to chaotic diffusion in phase space.

## 2. The integrable model

Motion in a smooth gravitational field becomes quite simple if the number of isolating integrals equals the number of degrees of freedom, and much work in galactic dynamics has focused on finding integrable models for galactic potentials. Kuzmin, (1956, 1973) showed that there is a unique, ellipsoidally stratified mass model for which the corresponding potential has three global integrals of the motion, quadratic in the velocities. Kuzmin's model, explored in detail by de Zeeuw (1985) who christened it the "Perfect Ellipsoid", has a large, constant–density core in which the orbital motion is simillar to that of a three-dimensional linear oscilator. Every orbit in the core of the Perfect Ellipsoid fills a region close to a rectangular parallelepiped, or box. These trajectories were called box orbits. At higher energies in the Perfect Ellipsoid, box orbits persist and three new orbital families appear: the tubes (inner and outer long axis tubes and short axis tubes), orbits that preserve the direction of their circulation around either the long or short axis of the figure. Tube orbits respect an integral of motion analogous to the angular momentum, and hence -unlike box orbits- avoid the center. As pointed out by Schwarzschild (1981), these four families, box, inner and outer long axis tubes and short axis tubes, reproduce the general form of triaxial galaxies. The Perfect Ellipsoid model has been used by serveral authors as a first approximation to represent an elliptical galaxy.

In this section, we address some of the most known relevant aspects of the integrable Stäckel model -separable in ellipsoidal coordinates- whose three global integrals of motion admit analytical expressions.

The Perfect Ellipsoid is represented by an stratified density function distributed over concentric ellipsoids of semi-axes $ma$, $mb$ and $mc$, given by

$$\rho = \frac{\rho_0}{(1+m^2)^2}, \qquad (1)$$

where $\rho_0$ represents the central density and $m$,

$$m^2 = \frac{x^2}{a^2} + \frac{y^2}{b^2} + \frac{z^2}{c^2}, \qquad a \geq b \geq c \geq 0, \qquad (2)$$

is constant on an ellipsoidal shell. Following Chandrasekhar (1969), it is possible to obtain the potential for the Perfect Ellipsoid, which in ellipsoidal coordinates $(\lambda, \mu, \nu)$ adopts the Stäckel form:

$$V(\lambda, \mu, \nu) = -\frac{1}{4h_\lambda^2} \frac{G(\lambda)}{(\lambda+\beta)} - \frac{1}{4h_\mu^2} \frac{G(\mu)}{(\mu+\beta)} - \frac{1}{4h_\nu^2} \frac{G(\nu)}{(\nu+\beta)}, \qquad (3)$$

where $\alpha = -a^2$, $\beta = -b^2$, $\gamma = -c^2$ and $h_\lambda^2$, $h_\mu^2$, $h_\nu^2$ are the metric coefficients of the ellipsoidal coordinates and $G(\tau)$ is a function given in terms of an elliptical integral of the third kind (see de Zeeuw, 1985).

The Stäckel model has three explicit global analytic integrals, namely, the adelphic integrals $I_2$ and $I_3$ besides the total energy $H$, and indeed all orbits in it can be determined in terms of simple quadratures.

The adelphic integrals can be considered as generalizations of the angular momentum integrals that exist in axisymmetric and spherical potentials, but also as generalizations of the energy integrals always present in separable potentials in Cartesian coordinates. Every orbit in an Stäckel potential is the sum of three motions, one in each coordinate. As a result, motion is bounded by coordinate surfaces.

The adelphic integrals $I_2$ and $I_3$ are, in fact, linear combinations of other integrals $J$ and $K$:

$$I_2 = \frac{\alpha^2 H + \alpha J + K}{\alpha - \gamma}, \qquad I_3 = \frac{\gamma^2 H + \gamma J + K}{\gamma - \alpha}, \qquad (4)$$

where the energy $H$, $J$ and $K$ are functions of the ellipsoidal coordinates and conjugate momenta (see de Zeeuw, 1985).

Also as a consequence of the separability of the Stäckel potential, can the equations of motion be recast as:

$$p_\tau^2 = \frac{H - V_{eff}(\tau)}{2(\tau + \beta)}, \qquad (5)$$

where $\tau = \lambda, \mu, \nu$, and $V_{eff}(\tau)$ denotes the effective potential,

$$V_{eff}(\tau) = \frac{I_2}{(\tau + \alpha)} + \frac{I_3}{(\tau + \gamma)} - G(\tau), \qquad (6)$$

This characteristic of the system allows to analyze the orbital structure according to the values of $p_\tau^2$ for each set of chosen values of the integrals. Since the motion is only possible for $p_\tau^2 \geq 0$, $\tau = \lambda, \mu, \nu$, the motion in each coordinate is either an oscillation between turning points defined by $p_\tau^2 = 0$, or a rotation whenever $p_\tau^2 > 0$ for all $\tau$. The combination of these two kinds of motion will determine the class of orbit. The ranges where $p_\tau^2$ are non-negative define a volume space, determined by the integrals, where the motion takes place. Thus, the orbital structure of the system may well be studied by analyzing the values of (5) for different combinations of chosen values for its integrals. Such a systematic study can be performed by fixing one of the integrals to compute the momenta in the space determined by the two remaining ones.

de Zeeuw (1985) carried out a detailed classification of orbits for this Stäckel model, by analizing the tridimensional space of integrals. A projection of this space on the energy surface is shown in Fig. 1, where the two inner curves in the figure separate the four different orbital families: box, inner and outer long axis tubes and short axis tubes. In other words, these curves are separatices of different kind of motion and they play a central role in the dynamics of the system when it is perturbed (see section 4).

Though the properties of this model have already been thoroughly studied (de Zeeuw and Lynden-Bell, 1985), we will complement them by computing and analyzing the resonance structure of the system in its integrals' space. The knowledge of this resonance structure will allow us to forecast the global dynamical behaviour of the system when a non–integrable perturbation is introduced.

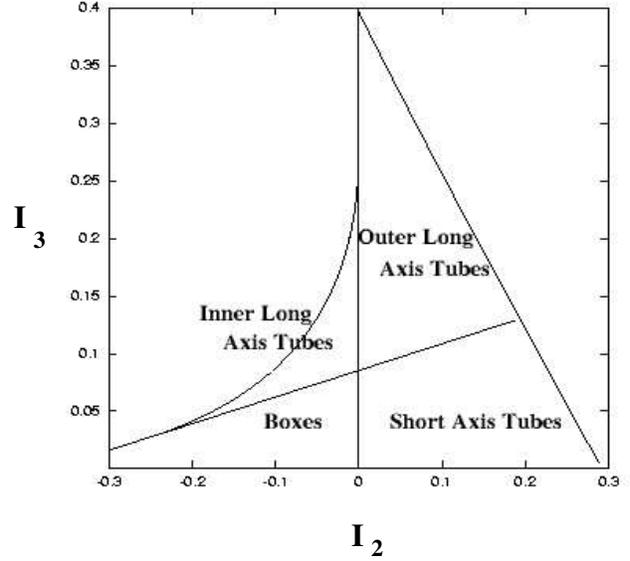

**Fig. 1.** Orbital Classification for the Perfect Ellipsoid in the $I_2 I_3$-plane, for a fixed value of the energy $E = -0.3$, de Zweeu (1985). Curves drawn in this plane are separatrices of different families of orbits.

## 3. Resonance Structure of the integrable model

In order to understand how an integrable system is modified by the effect of a non–integrable perturbation, it is worth-having a picture of its resonance structure, since it yields a preliminary knowledge of the way in which resonances in the integrable system may be distorted when a perturbation is turned on, avoiding the integration of the equations of motion and eventually the variational ones. Further, we can guess something about the mechanisms that could give rise to the transition from regularity to chaos by different processes.

In the present section, we compute the resonance structure for the integrable Stäckel model in the non-perturbed integrals space $\boldsymbol{I} = (H, I_2, I_3)$. At this stage we introduce action-angle variables, which in many respects are the natural variables for the description of the motion. Indeed, from the geometrical point of view the actions (which are functions of the global integrals) define the tori structure that foliate the phase space. For given values of the action, the angle variables describe the orbit on the torus.

Thus, we briefly outline the procedure for the case of motion separable in ellipsoidal coordinates, whose results applyes not only to the particular case of the Perfect Ellipsoid, but to every Stäckel potential. The action variables $J_\tau$ associated to each ellipsoidal coordinate are defined by

$$J_\tau = \frac{1}{2\pi} \oint p_\tau d\tau = \frac{2}{\pi} \int_{\tau_{min}}^{\tau_{max}} p_\tau d\tau, \qquad \tau = \lambda, \mu, \nu, \qquad (7)$$

where, from (5) and (6), the momenta $p_\tau$ can be written in the fashion (de Zweeu, 1985)

$$p_\tau^2 = \frac{H - I_2/(\tau + \alpha) - I_3/(\tau + \gamma) + G(\tau)}{2(\tau + \beta)}, \qquad (8)$$

and the integration is performed over all values of $\tau$ for which $p_\tau{}^2 \geq 0$.

From (7) and (8) we obtain $J_\tau = J_\tau(H, I_2, I_3)$ and denoting the action vector by $\boldsymbol{J} = (J_\lambda, J_\nu, J_\mu)$, we can formally invert this latter relation in order to get $H(\boldsymbol{J})$. For given values of the integrals, $H = E$, $I_2 = i_2$ and $I_3 = i_3$, we get the values of the actions $J_\tau$ that fix the torus where the motion takes place. The associated angle variables $\theta_\tau$, canonically conjugate to the actions $J_\tau$, are linear functions of the time $t$,

$$\theta_\tau = \omega_\tau(\boldsymbol{J}) \cdot t + \theta_\tau(0), \qquad (9)$$

where $\theta_\tau(0)$ are constants determined by the three remanent initial conditions and the quantities (integrals) $\omega_\tau(\boldsymbol{J})$ given by

$$\omega(\boldsymbol{J}) = \frac{\partial H(\boldsymbol{J})}{\partial \boldsymbol{J}} \qquad (10)$$

are the frequencies of the motion in each degree of freedom.

If all the three frequencies are incommensurable, the orbit fills densely the volume allowed to it by the values of the integrals of motion, i.e. the region where $p_\tau^2$ are non-negative. In other words, for these particular values of the actions (or the integrals) we have a strong non–resonant 3D torus on which the motion is ergodic. This occurs when the frequencies satisfy the so–called diophantine condition (see, for instance, Giorgili, 1990) which states that for a given integer vector $\boldsymbol{m}$ the frequency vector satisfies

$$|\boldsymbol{m} \cdot \boldsymbol{\omega}| \geq \gamma |\boldsymbol{m}|^{-\alpha}, \qquad |m| = |m_1| + |m_2| + |m_3|, \qquad (11)$$

where $\gamma > 0$ and $\alpha > 2$ in case of a 3D system. Whenever a relation of the form

$$\boldsymbol{m} \cdot \boldsymbol{\omega}(\boldsymbol{J}) = 0, \qquad \boldsymbol{m} \in \mathbb{Z}^3/\boldsymbol{0} \qquad (12)$$

is satisfied, we obtain a resonance condition for the actions or the frequencies. Those values of the actions that fulfill (12) lead to a resonant torus, that is, orbits are not ergodic on a 3D torus but on a 2–dimensional one. It may be interpreted that the resonance condition provides a relation between the actions that lead to a new *local* integral which confines the motion to a manifold of lower dimensionality.

In order to obtain the resonance structure on an energy surface, a particular value of the energy $H$ has to be fixed and the resonance condition (12) solved for each value of $\boldsymbol{J}$.

Let us recall that frequencies for ellipsoidal coordinates are given by (10) and, since actions do not admit explicit analytical expressions in terms of the integrals, frequencies (10) have to be calculated by numerical means. Indeed, since $H = H(J_\lambda, J_\mu, J_\nu)$, we may write (de Zweeu, 1985)

$$\begin{aligned}
\frac{\partial H}{\partial H} &= \omega_\lambda \frac{\partial J_\lambda}{\partial H} + \omega_\mu \frac{\partial J_\mu}{\partial H} + \omega_\nu \frac{\partial J_\nu}{\partial H} = 1, \\
\frac{\partial H}{\partial I_2} &= \omega_\lambda \frac{\partial J_\lambda}{\partial I_2} + \omega_\mu \frac{\partial J_\mu}{\partial I_2} + \omega_\nu \frac{\partial J_\nu}{\partial I_2} = 0, \\
\frac{\partial H}{\partial I_3} &= \omega_\lambda \frac{\partial J_\lambda}{\partial I_3} + \omega_\mu \frac{\partial J_\mu}{\partial I_3} + \omega_\nu \frac{\partial J_\nu}{\partial I_3} = 0.
\end{aligned} \qquad (13)$$

Then,

$$\omega_\lambda = \frac{1}{\Delta}\frac{\partial(J_\mu, J_\nu)}{\partial(I_2, I_3)}, \quad \omega_\mu = \frac{1}{\Delta}\frac{\partial(J_\nu, J_\lambda)}{\partial(I_2, I_3)}, \quad \omega_\nu = \frac{1}{\Delta}\frac{\partial(J_\lambda, J_\mu)}{\partial(I_2, I_3)}, \qquad (14)$$

where we have written

$$\Delta = \frac{\partial(J_\lambda, J_\mu, J_\nu)}{\partial(H, I_2, I_3)}. \qquad (15)$$

The partial derivatives that occur in the above expressions can be evaluated by differentiating under the integrals, (7)

$$\begin{aligned}
\frac{\partial J_\tau}{\partial H} &= \frac{2}{\pi} \int_{\tau_{min}}^{\tau_{max}} \frac{d\tau}{(\tau+\beta)p_\tau}, \\
\frac{\partial J_\tau}{\partial I_2} &= -\frac{2}{\pi} \int_{\tau_{min}}^{\tau_{max}} \frac{d\tau}{(\tau+\alpha)(\tau+\beta)p_\tau}, \\
\frac{\partial J_\tau}{\partial I_3} &= -\frac{2}{\pi} \int_{\tau_{min}}^{\tau_{max}} \frac{d\tau}{(\tau+\beta)(\tau+\gamma)p_\tau}.
\end{aligned} \qquad (16)$$

The values $\tau_{min}$ and $\tau_{max}$ define the range in each ellipsoidal coordinate where the motion is allowed, that is, the coordinate space regions where the condition $p_\tau^2 \geq 0$ is satisfied. The above formulae permit the calculation of the actions $J_\tau$ and the frequencies $\omega_\tau$ without integrating the equations of motion. For a Stäckel potential, and given values $E, i_2, i_3$ of the integrals of motion, the actions follows from (7), and the frequencies can be computed from (14) and (16).

Thus, in order to compute the frequencies for particular values of $H, I_2$ and $I_3$, we have to integrate (16) by numerical means and obtain the roots of $p_\tau^2 = 0$, which provides the values of $\tau_{min}$ and $\tau_{max}$. This numerical procedure restricts the accuracy with which the frequencies are obtained. In fact, we have calculated the integration limits in (16) by means of the *zbrent* subroutine (Press et al., 1994), setting the tolerance parameter equal to $10^{-6}$. Further, we have performed the numerical integration for (16), by means of the Romberg method -implemented in the *midpoint* and *qromo* subroutines (Press et al., 1994)- with an accuracy of order of $10^{-10}$. Also, it has turned out that for particular values of the integrals $H$, $I_2$ and $I_3$, the functions involved in (16) are not well-defined for $\tau_{min}$ and $\tau_{max}$. This shortcoming has been avoided by shrinking the integration ranges to $(\tau_{min} + \delta\tau, \tau_{max} - \delta\tau)$, being $d\tau \sim 10^{-4}$ a suitable value rising as a compromise between the need of minimizing the final errors in the frequencies and the computational time required to the whole numerical procedure.

From the analysis of all the above discussed factors that affect the accuracy with which the frequencies can be obtained, we do take in practice as the resonance condition the relation that follows

$$|\boldsymbol{m} \cdot \boldsymbol{\omega}(\boldsymbol{J})| \approx 10^{-4}. \qquad (17)$$

which determines the different resonant surfaces in the integrals'space $(H, I_2, I_3)$.

In order to visualize the resonance structure of the Stäckel model, let us fix a value for the energy $H$, which defines the energy surface in the integrals'space. Fig. 2 shows the resonance structure for the Perfect Ellipsoid in the $I_2I_3$-plane, corresponding to $H = -0.3$, being this particular value of the energy in the range $-0.9074 \leq E < 0$ corresponding to the Stäckel model, a representative one of the dynamical behavior of the system.

The curves shown in this figure arise from the intersections between the energy surface and several resonant surfaces calculated from (17), in terms of $I_2$ and $I_3$, for different resonant vectors $m$ satisfying the condition $|m| = |m_1| + |m_2| + |m_3| < 8$. Let us mention that, as far a we know, this is the first attempt to compute the resonance structure for this model.

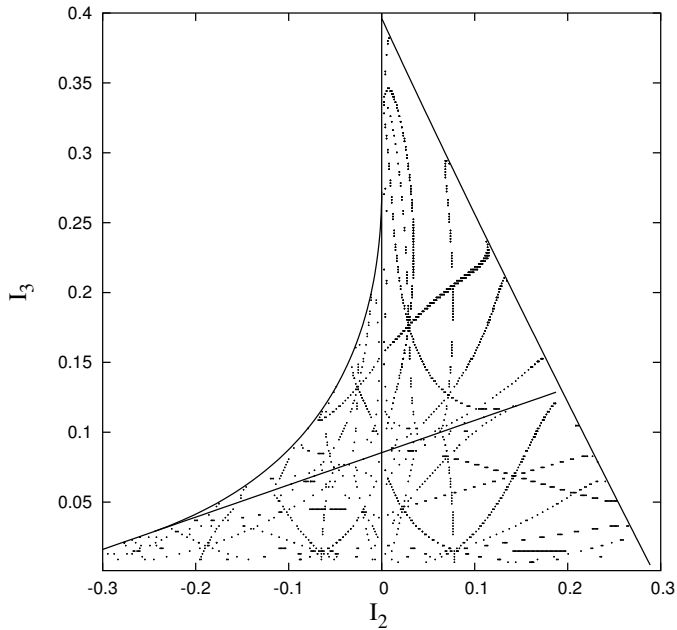

**Fig. 2.** Resonance structure of the Perfect Ellipsoid onto the $I_2I_3$-plane, for a particular value of the energy $E = -0.3$. The curves shown in this figure are the intersection of the energy surface and several resonant surfaces calculated from (17) for different resonant vectors $m$ satisfying $|m| = |m_1| + |m_2| + |m_3| < 8$ (see text).

In order to understand how the resonance structure of the Stäckel model determines the orbital structure of the system, in Fig. 3 we present three particular resonances, namely, $(1, -3, 2), (1, -2, 1)$ and $(3, -1, -1)$, as well as three tori located in different regions on the $I_2I_3$-plane. Over these tori, orbits will proceed with different dynamical features depending on their location with respect to the resonant curves. Thus, the non-resonant torus depicted in the figure lies in a region of the phase space where the resonant condition does not hold for any $m$ (or where the diophantine condition (11) holds), and any orbit lying on such a 3D torus will cover it densely and uniformly as $t \to \infty$. As mentioned above, in the 3D torus the motion is ergodic. Meanwhile, motion on a torus located on a resonant curve is such that any trajectory inhabits a submanifold of dimensionality two as a consequence of the resonance condition. Therefore, an orbit on such a torus will densely cover a 2D (elliptic) torus, the resonant torus. Closure in configuration space requires an additional, independent resonance condition. Any torus located on the intersection of two resonances will lead to a periodic orbit since the two independent resonance conditions restrict the motion to one dimension (a 1D torus).

It is worthwhile to taking a look at both Figs. 1 and 2 in order to have a description of the orbital structure in the Stäckel model. Let us remark that the internal curves in Fig. 1, given by $I_2 = 0$ and $E = V_{eff}(-\beta)$, play an important role as they bound as well as separate the four regions corresponding to the four families orbits for the Perfect Ellipsoid. The separatrices are very unstable, even a very small perturbation dramatically modifies the dynamics on them. Technically, in the integrable model the smooth separatrices are present because the stable and unstable manifolds of a given object (in this case a 2D hyperbolic torus) exactly match. Note, however, that in the surroundings of the separatrices, the resonance structure shows rather intricate.

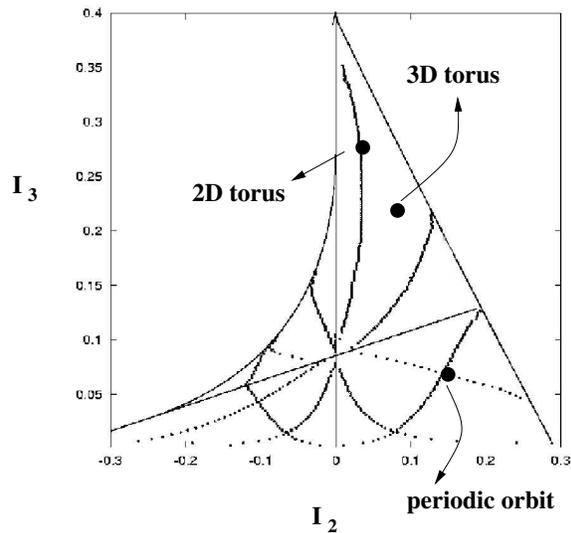

**Fig. 3.** The three different resonances $(1, -3, 2), (1, -2, 1)$ and $(3, -1, -1)$ showed in the figure were extracted from Fig. 2 (see text).

The resonance structure of the integrable Stäckel model allows us to forecast which the dynamical behavior of the system could be if a non–integrable perturbation would be added. In fact, as it will be discussed in the next seccion, the interaction of resonances are responsible for the arising of chaotic motion when a perturbation is turned on.

## 4. Theoretical discussion of the perturvative effects

Though a deep insight on the dynamics of the perturbed Stäckel system gained through numerical means will be addressed in a forthcoming paper, several of its main features can be inferred from the resonance structure of the integrable model.

We begin this section by reviewing a theoretical discussion in order to comprehend how a perturbation would modify the dynamics of the integrable Stäckel model, following the general description given by Chirikov (1979), Cincotta (2002). With this purpose, let us be concerned with a system represented by the Hamiltonian

$$H(J, \theta) = H_0(J) + \epsilon V(J, \theta), \quad (18)$$

where

$$\epsilon V(J, \theta) = \epsilon \sum_m V_m(J) \cos(m \cdot \theta). \quad (19)$$

Here $(J, \theta)$ denote the tridimensional action-angle coordinates for the unpertubed Hamiltonian $H_0$, $m$ is a tridimensional in-teger vector, $V_m$ certain real functions and $\epsilon$, the perturbation parameter, is a real number.

Elliptical galaxies might be represented by a smooth triaxial potential $H_0$, like the Stäckel one, plus a perturbation given by a multipolar expansion, $\epsilon V$. We are therefore assuming that the perturbation is analytic, which may not be the case if we add a central singularity.

For $\epsilon = 0$ we have $H = H_0(J)$ and the motion is absolutely stable[1] for any initial condition, since we have the complete set of the three integrals of motion $J_i$ ($H_0$ being cyclic in $\theta$), that is to say that the Hamiltonian is completely integrable. The phase vector evolves linearly with time with a frequency vector $\omega(J) = \partial H_0/\partial J$, with $det(\partial \omega_i/\partial J_j) \neq 0$ in order that $\omega(J)$ be a one-to-one function. This condition guarantees the nonlinearity of the system.

The presence of the perturbation in general disrupts the integrability of the Hamiltonian $H_0$ (due to the phase dependence of $V$), leading to a variation of the unperturbed integrals $J_i$. The stability of the motion breaks down when a large change in the actions takes place, i.e, when a 'gross' instability sets up.

To describe the motion of a star in the Hamiltonian (18)–(19), we take advantage of a perturbative technique in order to obtain approximate solutions, on assuming that the perturbation parameter is small, $\epsilon \ll 1$, that is, we consider a near-integrable Hamiltonian system. The perturbative approach given by the so-called asymptotic series implies, roughly, that the variation of the unperturbed actions is computed via a power series in the perturbation parameter.

It is well known that the effect of a perturbative Fourier component (like in (19)) is stronger as long as the time variation of its phase, $\dot\psi_m = m \cdot \dot\theta$, is slow. In the limit case of constant phase we reach the resonance condition for the unperturbed frequencies. If we are far from a resonance (i.e., the initial conditions correspond to values of the actions or the frequencies that satisfy the diophafntine condition (11)), it can be shown that the motion is stable. On taking advantage of the so-called averaging method, we neglect the oscillating part of the perturbation retaining only its average value.

When we are near to a resonance condition, the asymptotic series technique does not work any longer due to the appearance of the so-called small denominators in the coefficients of the asymptotic series. These small denominators are the resonant values $\omega_m = m \cdot \omega$ that may lead to divergent series. It can be shown that the set $\omega_m$ for all integer vectors is, in general, everywhere dense in phase space. Therefore, to find initial conditions 'far from a resonance' is not an obvious task.

The geometric features of resonances in the action or integral space are represented in Figs. 2 and 3. Any point on this plane is a torus since the 'position vector' is given by the three values of the actions (or the integrals). In action space, the resonance equation $m \cdot \omega(J) = 0$ leads to some other 2-dimensional surface, whose local normal at the resonant point $J = J^r$ is

$$n^r = (\partial[m \cdot \omega(J)]/\partial J)_{J^r}. \qquad (20)$$

[1] Except for the unstable periodic orbtis, which form a set of zero measure.

Further, the conservation of the unperturbed energy provides the 2-dimensional energy surface $H_0(J) = E$. In what follows we will consider only *convex* Hamiltonians, condition that is verified by the Stäckel Hamiltonian. The subspace defined by the intersection of both the resonant and the energy surfaces has dimension 1.

By definition, the frequency vector $\omega$ is normal to the energy surface in the action space. The latter condition, together with the resonance equation, shows that the resonant vector $m$ lies in the tangent plane to the energy surface at $J = J^r$ Furthermore, a simple inspection of the equations of motion for only one resonant perturbing term, that is $\epsilon V = \epsilon V_{m_g}(J) \cos(m_g \cdot \theta)$, shows that $\dot J$ is parallel to the resonant vector $m_g$. This graphic picture of the dynamics allows us to conclude that the motion under a single resonant perturbation proceeds on the tangent plane to the energy surface at the point $J = J^r$ along the direction of the resonant vector. We may say then that as $\epsilon \to 0$, the resonant perturbation preserves the unperturbed energy.

The above discussion considering only one resonant term shows that the perturbation (19) only depends on a single phase, namely, the *resonant phase* $m_g \cdot \theta$. If we perform a canonical transformation in such a way that one of the three new phases, say $\psi_1$, is such that $\psi_1 = m_g \cdot \theta$, then the resulting Hamiltonian will be cyclic in the remainder two phases $\psi_2, \psi_3$ and the new momenta $p_2, p_3$ will be integrals of the motion.

This canonical transformation may be done through

$$\psi_i = \mu_{ik}\theta_k; \qquad J_i = J_i^r + p_k\mu_{ki}, \qquad (21)$$

where the sum over repeated indexes is understood and $\mu_{ik}$ is a $3\times 3$ matrix. In fact, this linear transformation could be thougth as a local change of basis where the new basis vectors are:

$$\mu_1 = m_g, \qquad \mu_2 = \frac{\omega(J^r)}{|\omega(J^r)|}, \qquad \mu_3 \equiv e \qquad (22)$$

being the latter normal to both $\mu_2$ and to $n^r$. Let us note that in general $e$ will not be normal to $m_g$.

According to the discussion given by Chirikov (1979), Cincotta (2002), the Hamiltonian (18)-(19) in the vicinity of a single resonant perturbation reduces to a simple pendulum model:

$$H_r(p_1, \psi_1) \approx H_0(J^r) + \frac{p_1^2}{2M} + \epsilon V_{m_g}(J^r) \cos \psi_1, \qquad (23)$$

where $H_r$ stands for the *resonant Hamiltonian* and $M$ is the 'non-linear mass' given by: $1/M = m_{gi}(\partial \omega_i/\partial J_k)_{J^r} m_{gk}$. Thus $p_1$ changes with time following the pendulum model for a given value of the integral $H_r$ in the $m_g$ direction. The remaining momenta $p_k, k = 2, 3$ are set equal to zero so that $J^r$ is a point of the trajectory. Therefore, since $H_r$ is independent of time, we have again the full set of three (local) integrals of motion for the problem of a single resonant perturbation: $H_r, p_2, p_3$.

The variations of the unperturbed actions are determined by the strengh of the perturbation and the non-linear mass, in the fashion (see Cincotta, 2002 for further details)

$$(\Delta J)_r \equiv |J - J^r| \sim \sqrt{|V_{m_g}(J^r)M|\epsilon}. \qquad (24)$$

Thus, the resonant surface of the unperturbed system in action space becomes actually a *resonant layer*, whose width is given by $(\Delta J)_r$.

Since $\boldsymbol{J} - \boldsymbol{J}^r \equiv \boldsymbol{p}$ and in the new basis $\boldsymbol{p} = p_1\boldsymbol{\mu}_1 + p_2\boldsymbol{\mu}_2 + p_3\boldsymbol{\mu}_3$ we note that being $p_2$ and $p_3$ local integrals, the motion proceeds from the resonant point in the direction of $\boldsymbol{\mu}_1$, i.e. across the resonance.

Cincotta (2002), following Chirikov, (1979), discusses how different terms in the perturbation modify the dynamics of the integrable system $H_0$. For the sake of clarity let us recast (18)-(19) as

$$H(\boldsymbol{J}, \boldsymbol{\theta}) = H_0(\boldsymbol{J}) + \epsilon V_{\boldsymbol{m}_g}(\boldsymbol{J})\cos(\boldsymbol{m}_g \cdot \boldsymbol{\theta}) +$$
$$\epsilon V_{\boldsymbol{m}_l}(\boldsymbol{J})\cos(\boldsymbol{m}_l \cdot \boldsymbol{\theta}) + \epsilon \sum_{\boldsymbol{m} \neq \boldsymbol{m}_g, \boldsymbol{m}_l} V_{\boldsymbol{m}}(\boldsymbol{J})\cos(\boldsymbol{m} \cdot \boldsymbol{\theta}). \quad (25)$$

If we take into account every term in (25), all momenta change with time and then variations in $p_2$ and $p_3$ could occur, since now the Hamiltonian depends on $\theta_2$ and $\theta_3$. These variations may proceed along the $\boldsymbol{\mu}_2$ and $\boldsymbol{\mu}_3$ directions, and recalling that $\boldsymbol{\mu}_2$ is normal to the unpertrubed energy surface, the variations in $p_2$ take into account small changes in the unperturbed energy when the pertubation is switched on. These variations are bounded and of order $\epsilon$. However, $\boldsymbol{\mu}_3$ is tangent to the energy surface and by definition it is directed along the resonance, so one could expect variations in $p_3$ that, at first sight, will be undbounded (see next section).

Let us now discuss the effects of the different terms in (25). As shown above, on taking $V_{\boldsymbol{m}} = 0$ for all $\boldsymbol{m} \neq \boldsymbol{m}_g$ and picking up initial conditions such that $\boldsymbol{m}_g \cdot \boldsymbol{\omega}(\boldsymbol{J}^r) = 0$, the tori structure in the vicinity of $\boldsymbol{J}^r$ is preserved because of the existence of the three local integrals $H_r, p_2$ and $p_3$, being $H_r(p_1, \psi_1)$ the pendulum Hamiltonian. Depending on the value of $H_r$, the momenta $p_1$ will osscillate or rotate around $\boldsymbol{J}^r$. The resonant layer corresponds to the oscillation domain of the pendulum energy. As it is well–known osciations and rotations in a pendulum are separated by a smooth separatix $p_1^s(\psi_1^s)$.

If we switch on $V_{\boldsymbol{m}_l}$, which we assume to be the dominant non–resonant perturbing term, its main effect on the dynamics of the system in the vicinity of $\boldsymbol{J}^r$ would be to split the stable and unsable manifolds associated to the unstable points of the pendulum leading to a splitting of separatices and the smooth curve $p_1^s(\psi_1^s)$ becomes a stochastic layer of finite width. Thus, under the effect of this perturvative term, oscillations and rotations are actually separated by a layer of chaotic motion; $p_1(t)$ shows a bounded unstable, chaotic behavior confined to this thin layer. Fig. 4 illustrates this fact for two given resonances in the Stäckel model in the unperturbed integral space $(H_0, I_2, I_3)$ for a fixed value of $H_0$. There, the local basis $\{\boldsymbol{\mu}_1', \boldsymbol{\mu}_2', \boldsymbol{\mu}_3'\}$, at a given resonant torus $\boldsymbol{I}^r$, has been included, being the $\boldsymbol{\mu}_i'$ the $\boldsymbol{\mu}_i$ at $\boldsymbol{J}^r$ mapped to the integral space.

If we turn on the next relevant term in (25), say $V_{\boldsymbol{m}_d}$, it is possible to show that this third term might cause unbounded chaotic variations in $p_3$ while $p_1$ lies in the stochastic layer, there could be motion in the $\boldsymbol{\mu}_3$ direction, that is, along the stochastic layer of the resonance.

Finally, Fig. 5 illustrates in a schematical way how the whole resonant structure of the Stäckel model (given in Fig. 2) could be distorted by the effect of a generic perturbation. All resonaces would become layers, whose widths $(\Delta J)_r$ strongly depend on $\boldsymbol{J}^r, \boldsymbol{m}, V_{\boldsymbol{m}}$ but all of them having order $\epsilon$.

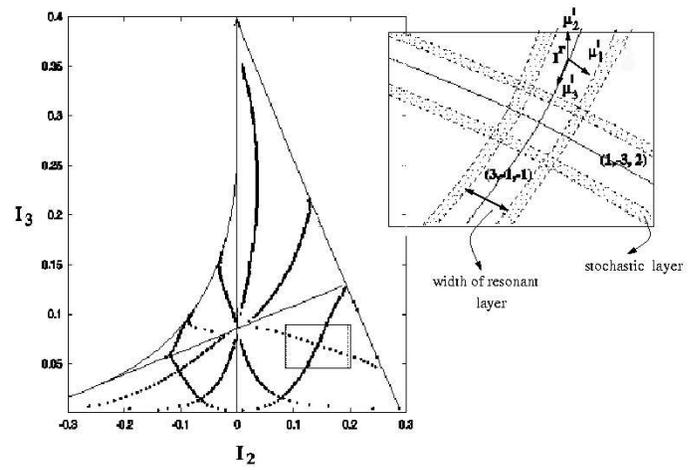

**Fig. 4.** The upper panel shows the resonances $(1, -3, 2)$ and $(3, -1, -1)$ of Fig. 3, with their concomitant resonant and stochastic layers.

Geometrically, when the system is exactly at some resonance, we have a 2D elliptical torus, while when the system is on the border of the resonance we get a 2D hyperbolic torus. Thus the center of a resonance layer over the integral space form a chain of elliptical tori while the border of the resonace layer corresponds to a chain of hyperbolic tori. This is schematically shown in Fig. 6 for the same resonances of the Stäckel potential depicted in Fig. 5.

All the orbits trapped in resonances constitute subfamilies of the Stäckel orbital families. As we discussed at the beginning of this section, if the system is far from any resonance, the motion is stable. The presence of the perturbation only produces small changes of order $\epsilon$ in the unperturbed actions. Thus a non-resonant value of the unperturbed actions or integrals leads to a 3D non-resonant torus that support quasiperiodic orbits. Accordingly to eqs. (7) and (8), the three local integrals are $H = H_0 + \epsilon V$, $I'_2 = I_2 + \epsilon \delta I_2$, $I'_3 = I_3 + \epsilon \delta I_3$, being $\delta I_i$ bounded constants. Therefore, in a neighbourhood of this values of the unperturbed integrals ($I_2, I_3$) the orbital structure is preserved.

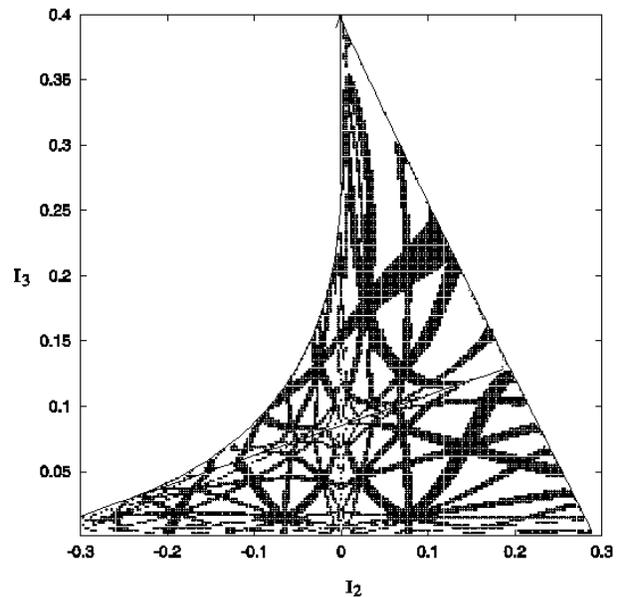

**Fig. 5.** Schematic representation of the effect of a perturbation on the resonant structure of the Stäckel model (given in Fig. 2). All the resonances become layers of width $(\Delta J)_r$.

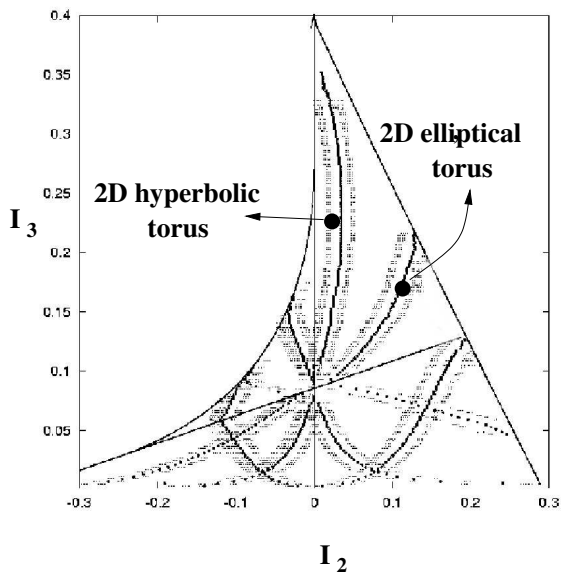

**Fig. 6.** Geometrical structure of the perturbed resonances shown in Fig. 5. The central curve, that corresponds to the exact resonance is a chain of 2D elliptical tori, while the borders of the resonance correspond to a chain of 2D hyperbolic tori.

On the other hand, orbits trapped in a single resonance at $J^r$, also respect three local integrals, that in action space are the pendulum Hamiltonian $H_r$ and two independent linear combinations of the unperturbed actions at the resonant value, for instance, $K_1 = a_1 J_1^r + a_2 J_2^r$, $K_3 = a_3 J_1^r + a_4 J_3^r$, where the coefficients $a_i$ depends only on the $\mu_{ki}$. For $H_r = 0$ we are exactly on a 2D elliptical torus, that supports a stable resonant orbit that is the same as in the unperturbed system (since $p_1 \equiv 0$ and $J = J^r$). But for positive values of $H_r$ new 3D tori appear that support nearly resonant orbits that respect the above mentioned local integrals. The nature of these new orbits are quite different from that of the unperturbed system at $J^r$. When $H_r$ takes the value that corresponds to the pendulum separatrix, we arrive to the 2D hyperbolic tori that support unstable orbits.

## 5. Discussion on the origin of chaos and diffusion

The dynamical description becomes largely intricate when we consider the interaction among all the perturvative terms. Indeed, as mentioned above, the effect of a perturbation on resonances is to form a thin stochastic layer at the border of the resonance layer. Besides, since the resonace layer have a width $\sim \sqrt{\epsilon}$, as long as the perturbation increases its strengh, the resonaces become wider and the overlap of the stochastic layers of different resonances coud take place when $\epsilon$ reaches some critical value $\epsilon_c$. When this occurs, we say that the resonances overlap. The dynamics in this case becomes largely unsatable due to the intersections of the stable and unstable manifolds of differ-ent resonances. The resulting motion is completely stochastic (see Cincotta 2002) whose main characteristic is its local instability. Fig. 5 provides a schematic illustration of the effect of a perturbation on the resonant structure. From this Fig. we can clearly see that if we increase the perturbation a massive overlap of resonaces will occur. Besides, due to the dimensionality of the system (3D), resonance intesections take place (which is not the case in 2D systems). Around the point of intersection, the stochastic layers of different resonaces ovelap, as Fig. 5 shows schematically.

For the time being, an analytical description of the dynamics at the resonance intersection is still lacking, because if we retain two resonant terms in the expansion (25), for instance $V_{m_g}$ and $V_{m_l}$, the Hamilonian depends then on two resonant phases $(m_g \cdot \theta)$ and $(m_l \cdot \theta)$, and as can be shown, this system is not integrable. A picture of the dynamics at a resonance crossing is given in Cincotta et al. (2003), where a zoom of the dynamical behaviour at the region close to the intersection of two resonances with their corresponding stochastic layers is presented. For this case, the intersection of these two resonances generates a stable periodic orbit and a central region of regular or mild chaotic motion bounded by a thin layer of unstable motion that presents at the same time several stable domains separated by extremely thin chaotic filaments. However, it is possible to infer that at the intersection of the unperturbed resonances a periodic orbit appears whose stability can not be determined a priori and that the zone of stochastic motion should be bounded (see Chirikov, 1979).

The kown mechanisms that lead to a transition from regularity to chaos are overlap of resonances (including resonace crossings) and Arnol'd diffusion–like processes (see for instance, Chirikov 1979, Cincotta 2002, Giordano and Cincotta 2004).

As we have already stated, chaos means variation of the unperturbed integrals. This is usually called in the literature chaotic diffusion. Unfortunatelly it does not yet exist any theory that could describe global diffusion in phase space. In other words, it is not possible to estimate neither its rate nor its direction or route. Though one could get accurate values of the Lyapunov exponents, the KS entropy, the MEGNO (see Cincotta et al. 2003, Cincotta and Simó 2000) or any other indicator of the stability of the motion, they provide only local values of the diffusion rate. A given orbit in a chaotic component of the phase space coud have for instance two positive and large values for the Lyapunov exponents which does not mean that the unperturbed integrals would change too much. This is a natural consequence of the structure of the phase space of almost all actual dynamcal systems like galaxies.

Thus what is actually relevant is the extent of the domain and the time–scale over which the diffusion may occur. In a recent work Giordano and Cincotta (2004) numerically show that in models similar to that of an elliptical galaxy, the time–scale over which diffusion becomes relevant is several orders of magnitude the Hubble time. On the other hand, in models corresponding to planetary or asteroidal dynamics, diffusion may occur over physical time–scales.

Anyway, one can make predictions about the global dynamical behaviour of a perturbed system by studying its resonance structure. For example, in Fig. 2 we can observe a rather intricate resonance structure in the box orbits domain, with several crosses of resonance as well as the fact that they are very close to each other. Thus, we expect that this region would be immediately dominated by a chaotic dynamics when the perturbation is activated, for instance by adding a mass concentration in the center of the system, the box domain in the $I_2I_3$–plane will be the first to be dominated by chaos. This is also true for any other kind of integrable model and perturbation, whatever its strengh at the center may be (see, for instance Papaphilippou, Y. and Laskar, J., 1998).

We could come to a similar conclusion for those regions close to the separatrices which separate different families of orbits, $I_2 = 0$ and $E = V_{eff}(-\beta)$. Several resonance intersections along these separatices can be observed, then, when a non–integrable perturbation is added, the dynamics in this region would become rather intricate due to the fact that the stable and unstable manifolds of different 2D hyperbolic tori will tangle in a very complicate fashion (see, for instance, Cincotta et al. 2003). This explains the instabilities found by de Zeeuw (1985) along the $I_3$-axis; in fact, in the Stäckel model all 2D tori with $I_2 = 0$ are unstable (hyperbolic), so that the dynamics in this region becomes rather chaotic under the effect of a non–integrable perturbation.

An important fact to be stated is that when chaos sets up, the unperturbed global integrals (or actions) have a discontinuos dependence on phase space variables. Indeed, close to the resonant torus, despite the existence of three local integrals, the unperturbed orbital structure is not preserved and the topology of the phase space changes. Moreover, on the stochastic layer at least one integral does not exist.

Thus, in the Stäckel model, for $\epsilon = 0$ we get Fig. 2, while for $\epsilon \ll \epsilon_c$ the dynamcial picture will be similar to that given in Fig. 5. Resonances do not present a significant overlap and chaotic motion will be confined to the narrow stochastic layers and around the ressonance crossings. Thus we cannot expect large variations of the unperturbed integrals and therefore chaotic diffusion should be irrelevant. However, theoretically, it might be possible that chaotic diffusion takes place even for very small values of the perturbation ($\epsilon \ll \epsilon_c$). Since for very small $\epsilon$ values the resonance structure is still present, one could expect that the perturvative terms drive diffusion along the stochastic layers of a given resonance, i.e. in $\mu'_3$ direction.

Indeed, Arnol'd, (1964) proved for a very specific Hamiltonian, that there exists a mechanism that could connect two distant tori in the chain of 2D hyperbolic tori located on the border of the resonance. This means that variations of the unperturbed actions along the resonance occur. Since the actual motion around any hyperbolic tori is stochastic, chaotic diffusion along the stochastic layer of a given resonance could take place. This mechanism that allows motion in the $\mu_3$ direction is the so–called *Arnol'd Mechanism* (see Giorgilli, 1990). However, the existence of a transition chain of hyperbolic tori does not mean that stochastic motion would spread over the whole web of resonances. Moreover, in general, it is not possible to find this transition chain in most non–linear Hailtonian systems, even in the case of the simpler ones. Nevertheless, it is often assumed that this is the case and that chaotic motion will take place over the whole resonance web. This implies that a global instability exists and it is usually called Arnol'd diffusion. This conjecture supports for instance, concepts like mixing or ergodicity among others. Let us recall that the global instability properties of near integrable Hamiltonian systems are far from being well–understood after the pioneer work of Arnol'd (see Lochak, 1999).

For $\epsilon \gtrsim \epsilon_c$, the overlap of resonances begins to operate destroying the tori structure of phase space and the resonance web disappears, and chaotic diffusion sets up. In this case, we expect that diffusion could operate through overlap of resonances, but the extent of the diffusive zone will be confined, at most, to the overlapping domain. There is no guarantee that the whole chaotic component of phase space is connected, even in the case of massive overlap (Giordano and Cincotta 2004). But also in this case, it is usually stated that the chaotic component of phase space is fully connected through Arnol'd diffusion. Arguments based on all these yet unproved statements have given rise to many of the current ideas taken for granted on dynamics in elliptical galaxies, as can be verified for instance in the works of Merritt and Valluri (1996), Merritt (1999), Udy, S. and Penniger (1988) among others.

Therefore, from the theoretical approach, though it is posible to get much information about the dynamical behaviour of the system as long as the perturbation increases its strenght, nothing can be concluded about the diffusive processes in phase space of near–integrable systems. Numerical integrations become the only way to investigate if chaotic diffusion could play a significant role in particular models of galaxies. This will be addressed in a forthcoming paper using the Stäckel potential here discussed.

Summing up, we would like to stress that we clearly discriminate the dynamical behaviour of the system depending on the values of the integrals. Close to strong non–resonant tori, the orbital structure of the unperturbed Stäckel potential is preserved and the local integrals are just corrections of order $\epsilon$ of the unperturbed global integrals. On the other hand, when the system is close to a resonant tori, the unperturbed orbital structure is not preserved, new local integrals appear and the topology of phase space changes. These new local integrals are the pendulum Hamiltonan and linear combinations of the unperturbed actions at the resonant point.

Therefore, in such a system, it is not possible to assume that the distribution function of the galaxy, in the whole regular component, has the form $f(I_2, I_3)$. This could be true only for strong non–resonant tori, but since resonances are dense in phase space one should state that the distribution function locally exists and has the form $f_n(H, I_2, I_3) + \epsilon g(H, I_2, I_3)$ in a neighbourhood of non–resonant tori that support quasiperiodic orbits and $f_r(H_r, K_2, K_3)$ in a vicinity of resonant tori.

On the other hand, nothing could be said about the dependence of $f$ in the chaotic region. In a similar fashion, since there is no theoretical support to argue that the whole chaotic region is fully connected, one should define a local distribution function only in those regions of phase space that could be visited by a single orbit. It is very likely that in some regions two local quasi-invariant exist (for example in the stochastic layers of the resonances) and in some other regions only the energy remains constant. Clearly, a notorious discontinuous dependence of $f$ on the integrals, and consequently on phase space variables, is expected.

On the other hand, much progress should be done in the understanding of diffusive processes in phase space of near–integrable Hamiltonian systems, including the dynamical behaviour around resonance crossings. Indeed, as all the evidence shows, galaxies should exhibit a divided phase space, with both regular and chaotic components, which is the phase space structure of near–integrable Hamiltonians for low–to–moderate perturbations.

Taking into account all this arguments, that are essentially supported on theoretical considerations, we suggest that this is the line of work in which we have to direct all our efforts in order to advance in the task of constructing equilibrium models of elliptical galaxies.

Let us conclude addresing some considerations about the analytical tools used along this paper.

The selected representation of an elliptical galaxy does not encompass the case of a central singularity. In such a case, one deals with a non–analytic perturbation which admits no Fourier expansion as (18) and the perturbative approach is no suitable. Nevertheless, it is clear that such a peturbation $\sim 1/r$ mainly affects the dynamics in the box domain due to the fact that these orbits pass arbitrarily close to the center: box orbits are strongly scattered producing in general an almost spherical configuration near the center (see however Poon and Merritt, 2002). Therefore, the perturbative approach here addressed should be reformulated in order that a singularity at the origin could be taken into account. This is not a simple task. Though there exists observational evidence of central mass concentrations in elliptical galaxies that could be black holes (see for instance Ferrarese and Ford, 2005), the question if they actually are supermassive black holes is still an open matter.

Due to the fact that the described perturbative approach succeeds in studying actual physical problems such as planetary and asteroidal systems or dynamics in particle accelerators as well as the motion of charged particles in magnetic bottles, we argue that the here discussed theoretical formulation, which on the other hand is yet the only one developed, is actually a fairly useful tool that should be seriously considered when dealing with galactic systems.

Though one may argue that the action-angular variables do not constitute an adequate set to yield a description of galatic systems, it can not be dennied that there exists wide numerical evidence of galatic models exhibiting large regions of phase space corresponding to regular moton. Therefore, the Hamiltonian of a star moving in a galactic potential would always allow to be written in the fashion $H(\boldsymbol{J}) = H_0(\boldsymbol{J}) + \epsilon V(\boldsymbol{J}, \boldsymbol{\theta})$, being the key point what should be considered as $V(\boldsymbol{J}, \boldsymbol{\theta})$. In fact, since all these mathematical theories have well served to provide a thorough description of chaotic systems, it seems to be no reason why they should not be of use to yield a good first order approximation to the motion of a star in a smooth stationary potential. Moreover, despite tha fact that the actual dynamics of a galactic system outcomes from numerical simulations, it is worth–counting with a theory that permits to elucidate the matter, at least in an heuristic way.

*Acknowledgements.* This work was partially supported by grants of the Consejo Nacional de Investigaciones Científicas y Técnicas (CONICET) and Fundación Antorchas. The authors aknowledge the valuable comments and suggestions of the refereee that served to enrichen the present effort.